# Bulk-Free Topological Insulator $Bi_2Se_3$ nanoribbons with Magnetotransport Signatures of Dirac Surface States

Gunta Kunakova[1,2], Luca Galletti[1], Sophie Charpentier[1], Jana Andzane[2], Donats Erts[2], François Léonard[3], Catalin D. Spataru[3], Thilo Bauch[1], and Floriana Lombardi[1]

[1] *Quantum Device Physics Laboratory, Department of Microtechnology and Nanoscience, Chalmers University of Technology, SE-41296 Gothenburg, Sweden*
[2] *Institute of Chemical Physics, University of Latvia, Raina Blvd. 19, LV-1586, Riga, Latvia*
[3] *Sandia National Laboratories, Livermore, CA, 94551, United States*

**Many applications for topological insulators (TIs) as well as new phenomena require devices with reduced dimensions. While much progress has been made to realize thin films of TIs with low bulk carrier density, nanostructures have not yet been reported with similar properties, despite the fact that size confinement should help reduce contributions from bulk carriers. Here we demonstrate that $Bi_2Se_3$ nanoribbons, grown by a simple catalyst-free physical-vapour deposition, have inherently low bulk carrier densities, and can be further made bulk-free by size confinement, thus revealing the high mobility topological surface states. Magneto transport and Hall conductance measurements, in single nanoribbons, show that at thicknesses below 30 nm the bulk transport is completely suppressed which is supported by self-consistent band-bending calculations. The results highlight the importance of material growth and geometrical confinement to properly exploit the unique properties of the topological surface states.**

Three-dimensional topological insulators (3D TIs) form a new class of quantum matter with an insulating bulk and conducting Dirac surface states topologically protected against time-reversal invariant perturbations. Spin momentum locking of the Dirac electrons opens up a variety of novel electronic phenomena and possible applications [1–5].

While much progress has been made in controlling the properties of thin films of 3D TIs, for applications it is necessary to generate structures with reduced dimensions, not only for the eventual high density of devices that is needed in most applications, but also because new phenomena are expected to arise at reduced dimensionality. TI nanoribbons are indeed the basic building block to design mesoscopic devices (e.g., quantum dots, quantum point contacts) that are promising in both fundamental research to explore confined topological modes [6–10] as well as for spintronics [11,12] and quantum information applications [13–16]. Quite recently, various theoretical proposal have shown the advantage of TI nanowires, with suppressed bulk conduction, to realize Majorana fermions [17], instrumental for topological quantum computation.

A key challenge for TIs has been the presence of residual bulk doping which has made it difficult to directly probe the Topological Surface States (TSS). Native defects in TIs, such as Se vacancies in $Bi_2Se_3$, act as charge donors giving rise to bulk carriers with density up to $10^{19}$ cm$^{-3}$ [18]. For single crystal materials great progress has been made to address this issue by compensating the defects with intentional substitutions [19]. In thin film the use of capping layers in combination with structurally matched buffer layers [20,21] have allowed to simultaneously suppress both interfacial and bulk defects, yielding very high mobilities. Unfortunately this approach has not been successful for lower dimensionality TI structures [22].

Here we show that the bulk doping problem in nanostructures can be solved through the growth of TI nanoribbons through a catalyst-free Physical Vapor Deposition (PVD) without requiring any additional intentional doping or capping layers. The nanoribbons have low intrinsic bulk carrier densities and yield an exceptional high mobility of the topological surface state.

By combining Shubnikov-de Haas (SdH) oscillations and Hall effect measurements on $Bi_2Se_3$ nanoribbons, as a function of thickness, we discover a regime where the bulk carriers are fully suppressed. We demonstrate that the electronic transport in our nanoribbons has three main contributions: 1) Dirac electrons coming from the TSS at the top interface with the vacuum, 2) bulk carriers, and 3) carriers due to a high-density accumulation layer at the interface with the substrate. The nanoribbons become bulk-free, by size depletion of carriers for thicknesses below 30 nm.

Free-standing bismuth chalcogenide nanoribbons $Bi_2Se_3$ were grown on a glass substrate using a catalyst-free PVD method [23]. Nanoribbons obtained with this method have a thickness between 10 – 80 nm, and width ranging between 50 to 450 nm; the length can be up to 30 $\mu$m. The nanoribbons were mechanically transferred to an n-type doped Si substrate with a 300 nm thick layer of $SiO_2$. The electrodes were defined using electron-beam lithography followed by evaporation of a gold layer. Prior to the deposition of 80 nm of gold, the samples were etched with $Ar^+$ ions and 3 nm layer of Ti or Pt was evaporated to achieve ohmic contacts.

Fig. 1a shows a scanning electron microscope image of a typical device with a Hall bar electrode geometry. In this configuration, the measurement of the transversal resistance $R_{xy} = V_{xy}/I$ (see Fig. 1a) as a function of the external magnetic field perpendicular to the nanoribbon axis, allows to calculate the 3D carrier density $n_{3D,H}$. The Hall coefficient $R_H$ is given by

$$R_H = t \frac{dR_{xy}}{dB} \times \frac{w}{w_H} = \frac{1}{n_{3D,H} e}, \qquad (1)$$

where we have considered a correction factor to account for the actual width of the nanoribbon $w$ and the distance between the transversal contacts $w_H$ [24]. In Eq. (1) $e$ is the elementary charge and $t$ is the nanoribbon thickness. The value of $R_H$, measured in a magnetic field range of ± 14 T, always shows a negative slope, indicating n – type carriers.

Fig. 1b shows the temperature dependence of the nanoribbon sheet resistance calculated as $R_{xx,sh.} = R_{xx} \times w/L$, where $L$ is the distance between the longitudinal contacts. Nanoribbons with thicknesses above 30 nm show a pronounced hump of the $R_{xx,sh.}$ with a maximum at around 220 K. Below this temperature the resistance decreases with the temperature reaching a saturation at about 2 – 10 K. We have previously observed a similar behaviour on both $Bi_2Se_3$ and $Bi_2Te_3$ nanoribbons [23]. The presence of a hump in $R_{xx,n}(T)$ can indicate a multiband transport[19,25].

The thickness dependence of $n_{3D,H}$, shown in Fig. 1c is calculated (using Eq. 1) from the $R_{xy}(B)$ slope at 0 – 2 T at T= 2K, for nanoribbons of different thicknesses; data points related to $n_{3D,H}$ extracted at higher fields 12-14 T are also reported for three nanoribbons with thicknesses covering the entire explored thickness range. The two values differ by less than 20%, which indicates that the main features of the $n_{3D,H}$ vs. thickness dependence are not affected by the range of magnetic field used to fit the Hall resistance.

A striking feature of the dependence shown in Fig. 1c is that $n_{3D,H}$ increases with decreasing the nanoribbon thickness. A similar dependence has been reported for thin films of $Bi_2Se_3$ [25] and attributed to the increased density of Se vacancies for thinner films. In our case the data in Fig. 1c are from the same synthesis batch that gives nanoribbons with different thicknesses. We therefore expect to have the same bulk doping for all thicknesses, which rules out Se vacancies as a possible cause of such dependence. The increase of $n_{3D,H}$ for $t < 30$ nm is instead indicative of a stronger contribution of surface carriers. In a scenario where surface carriers are formed, with electron density larger than the bulk, by reducing the nanoribbon thickness increases the effective carrier density since the bulk contribution becomes less dominant. (Additional effects due to band-bending will be discussed below).

To confirm this picture, we used additional independent measurements to extract the carrier densities of the bulk, $n_B$, and those of the surface states (topological and/or trivial). In what follows we will indicate the 2D carrier density at the nanoribbon top surface (which interfaces the vacuum), as $n_{TS}$, and that at the bottom in contact with the substrate as $n_{Int}$. To determine the various carrier concentrations, we measure the longitudinal $R_{xx}$ and transversal $R_{xy}$ magneto resistance. We first consider thin nanoribbons ($t \leq 30$ nm) where we assume a negligible bulk carrier contribution.

Fig. 2a shows the $R_{xx}$ as a function of the magnetic field for a 30 nm $Bi_2Se_3$ nanoribbon (device B13-E5). We observe typical Shubnikov-de Haas (SdH) oscillations. The frequency $F$ of the oscillations is given by the Onsager relation:

$$F = \left(\frac{\hbar}{2\pi e}\right) A_0, \qquad (2)$$

where $A_0$ is the cross-section of the Fermi surface. For Dirac fermions, $A_0 = \pi k_F^2$ and $n_{2D,SdH} = k_F^2/4\pi$. The Fourier transform of the oscillatory part of $R_{xx}$ with removed polynomial background shows a single dominating frequency $F = 99$ T, which according to Eq. (2) gives a 2D carrier density $n_{2D,SdH} = 2.4 \times 10^{12}$ cm$^{-2}$ with a corresponding mobility of 6800 cm$^2$(Vs)$^{-1}$ extracted from a Dingle analysis, (see Suppl. Info., Table S1). We have obtained similar results for two other nanoribbons with thicknesses ≤ 30 nm (Suppl. Info. Fig. S1, Table S1). The intercept extracted from the Landau level diagrams for these nanoribbons is close to 0.5 as expected for Dirac fermions possessing a Berry phase $\varphi_B = 2\pi\beta = \pi$ [26]. Because of the rather low carrier concentration, the highest magnetic field in our experiment (B = 14T) populates low order Landau levels (close to n = 0), which makes the extraction of the Berry phase quite reliable (see Suppl. Info. S3).

The angular dependence measured for this thin $Bi_2Se_3$ nanoribbon shows a typical $F \sim 1/\cos(\theta)$ dependence, where $\theta$ is the angle between the magnetic field direction and the surface of the nanoribbon. This confirms the 2D nature of the SdH oscillations (see Suppl. Info., Fig. S4) and allows us to conclude that the observed SdH oscillations in nanoribbons with $t \leq 30$ nm originate from a single TSS. We assign these TSS to the top surface (at the interface with vacuum) of the nanoribbon. This is because the SdH oscillations, for all nanoribbons, are not affected by a gate voltage while the Hall conductance can be tuned by a bottom gate (Suppl. Info., Fig. S9). We also exclude that the TSS at the bottom interface (in contact with the substrate) would contribute to the frequency $F$ we extract from the SdH oscillations, since this would lead to a modulation of $F$ by an applied gate voltage. The fact that the TSS at the bottom surface do not generate SdH oscillations up to 14 T could be related to its overlapping with the charge accumulation layer formed at the interface with the substrate, as we will show below.

The values of the 2D carrier densities extracted from the SdH oscillations and from the Hall effect measurements show a remarkable discrepancy. For example, for nanoribbon B13-E5 ($t$ = 30 nm) $n_{2D,SdH}$ is $2.4 \times 10^{12}$ cm$^{-2}$ while $n_{2D,H}$, calculated as $n_{3D,H} \times t$, is $1.7 \times 10^{13}$ cm$^{-2}$, almost an order of magnitude higher. The change of slope in the transversal resistance data *vs.* magnetic field shown in the inset of Fig. 1c is an indication of a contribution of a second band to the total carrier concentration. To extract the carrier concentration and the mobility of the second band (we neglect the bulk) we have performed a two-carrier analysis of the longitudinal and transversal magneto conductance measurements. The conductance tensor in the two-carrier analysis is described as [27]:

$$G_{xx}(B) = e \left( \frac{n_1 \mu_1}{1+\mu_1^2 B^2} + \frac{n_2 \mu_2}{1+\mu_2^2 B^2} \right) \tag{3}$$

$$G_{xy}(B) = eB \left( \frac{n_1 \mu_1^2}{1+\mu_1^2 B^2} + \frac{n_2 \mu_2^2}{1+\mu_2^2 B^2} \right). \tag{4}$$

Here $n_{1;2}$ and $\mu_{1;2}$ are the carrier density and the mobility of the conduction bands 1 and 2, respectively. The conductance tensor was calculated from the transverse and longitudinal resistances:

$$G_{xx} = \frac{R'_{xx}}{{R'_{xy}}^2 + {R'_{xx}}^2}; \; G_{xy} = -\frac{R'_{xy}}{{R'_{xy}}^2 + {R'_{xx}}^2}, \tag{5}$$

where $R'_{xx} = R_{xx} \times w/L_{xx}$ and $R'_{xy} = R_{xy} \times w/w_H$, and $R_{xy}$ is the measured transversal resistance, $w$ and $L_{xx}$ is the width and length of the nanoribbon, and $w_H$ is the distance between the Hall electrodes. To achieve a sufficient fitting accuracy within this model, we have included the correction for a narrow channel, $w/w_H$, considering the actual geometries of the nanoribbons and fixing the carrier density $n_2 = n_{2D,SdH}$ extracted from the SdH measurements. Fig. 2b shows $G_{xx}$ and $G_{xy}$ with the fit of the two-carrier model. The agreement is quite remarkable. The extracted parameters for the nanoribbon B13-E5 are $n_1 = 1.5 \times 10^{13}$ cm$^{-2}$, and $\mu_1$ and $\mu_2$ of 2930 and 10600 cm$^2$(Vs)$^{-1}$ respectively. We assign the $n_1$, extracted from the two-carrier model, to a carrier density deriving from the accumulation layer at the bottom surface; we assume therefore that $n_1 = n_{Int}$. The $n_1$ we extract from fitting (Eqns. 3 and 4) is only slightly lower than the experimental $n_{2D,H} = 1.7 \times 10^{13}$ cm$^{-3}$ indicating the accumulation layer dominates the Hall conductance. The discrepancy between $\mu_2$ extracted from SdH and the value obtained by the two-carrier analysis is most probably due to the different scattering times involved in Hall and SdH measurements [28,29].

In our two-carrier analysis, we have considered that the TSS at the bottom surface has a carrier mobility similar to the one of the trivial accumulation layer allowing us to describe both bands by a single carrier concentration $n_{Int}$. This assumption accounts for: a) the very good fitting of the conductance tensors without the addition of a third band and b) the fact that the TSS at the bottom have a much lower mobility. Indeed, in the case of comparable mobility between the TSS at the top and bottom surface one would expect to detect clear signatures in the SdH oscillation that we do not observe. Finally, the relatively lower mobility of the carrier bands at the interface with the substrate (accumulation layer plus TSS having similar mobilities) makes the condition $\mu B >> 1$, to observe quantum oscillation, not fulfilled at the fields used in our experiment (up to 14 T). The low carrier mobility of the carrier band at the interface with the substrate could possibly be related to an increased electron-electron scattering due to the larger carrier concentration compared to the top surface.

The Hall conductance gives $n_{3D,H} = 4 \times 10^{18}$ cm$^{-3}$ for thicknesses $t > 40$ nm (see Fig. 1c). We can therefore consider this value as the upper limit for the hypothetical bulk carrier concentration $n_B$ (that we have up to now neglected) of the thin nanoribbons. The Fermi energy for these hypothetical bulk electrons, measured from the bottom of the conduction band, can be calculated as [30] $E_F^B = \hbar^2/(2m^*)(3\pi^2 n_{3D,Bulk})^{2/3}$. Assuming that the effective mass of Bi$_2$Se$_3$ bulk carriers is 0.15 $m_e$ [31], $E_F^B$ is 60 meV. From the SdH oscillations one can extract the Fermi energy of the top surface $E_F^{TS}$ measured from the Dirac node: $E_F^{TS} = \hbar k_F v_F$, where $k_F$ is obtained from Eq. (2) and $v_F$ is taken to be equal to $5\times10^5$ m/s [32,33]. Based on this analysis, $E_F^{TS}$ is located in the bulk gap or at the minimum of the conduction band (see Table 1) demonstrating the effectiveness of our catalyst-free PVD method to grow nanoribbons with low residual doping. Taking into account that the Dirac point is about 180 meV below the bottom of the conduction band, one gets a bending of the minimum of the conduction band between the bulk and the top surfaces $\Delta E_{BB}^{TS} = E_F^B - (E_F^{TS} - 180$ meV$)$ [33,34].

In Table 1 we list the calculated $\Delta E_{BB}^{TS}$ energies for several Bi$_2$Se$_3$ nanoribbons. For all nanoribbons thinner than 30 nm the determined BB energy between the TTS and the hypothetical bulk is $\Delta E_{BB}^{TS} \approx +65$ meV indicating an upward band-bending associated with a depletion layer with depth $z_{Top}$. To confirm this picture and to evaluate the extension of the depletion $z_{Top}$ and accumulation $z_{Int}$ layers at the two interfaces, we performed self-consistent simulations of the band-bending as discussed in Methods using the boundary conditions taken from the experiment. Fig. 3a shows the evolution of the conduction band minimum (CBM) as a function of the distance from the substrate for 5 different nanoribbon thicknesses. In the plot, the Fermi level is at zero energy. The values of $z_{Top}$ and $z_{Int}$ are comparable and on the order of 15 nm; this implies that for nanoribbons less than 30 nm in thickness the carrier density is determined primarily by the surfaces instead of the bulk. We quantified this assessment by calculating the effective carrier concentration as a function of nanoribbon thickness, as shown in Fig. 3c. The results show the same qualitative behaviour as the experimental data of Fig. 1c.

To experimentally probe bulk carriers in thick nanoribbons we measured the Hall conductance and SdH oscillations of nanoribbons with thicknesses above 30 nm. The $R_{xx}$ as a function of $1/B$ clearly shows a multifrequency pattern. In Fig. 4a the SdH oscillations of a nanoribbon with thickness of 63 nm (device BR3-10R2) are shown at two different gate voltages. As one can see, the FFT spectra of $R_{xx}$ with subtracted background gives two dominating frequencies $F_1 = 45$ T and $F_2$ = 105 T, the latter being very close to the single-frequency SdH oscillation observed in the thin nanoribbons (Fig. 2a, inset and Suppl. Info. Fig. S1). At large gate voltages the charge carriers from the interface states can be tuned and the SdH oscillation frequency, corresponding to these carriers, should move with applied gate voltage [35]. Applying – 75 V to the bottom gate gives no change in the frequencies $F_1$ and $F_2$, indicating that $F_1$ and $F_2$ represent either the bulk or the TSS carriers at the top surface. These measurements further confirm that the interface carriers do not show up in the SdH oscillations, while contributing to the Hall conductance.

The 3D carrier density for the thick nanoribbons can be estimated from the relation $n_{3D} = 1/(2\pi)^2 (4/3) k_F^3$, where $k_F$ is the Fermi wave vector extracted from Eq. (2). The estimated $n_{3D}$ for the frequencies $F_1$ and $F_2$ are $1.7 \times 10^{18}$ and $6.1 \times 10^{18}$ cm$^{-3}$, respectively. $F_2$ gives a carrier density above what we have identified as being the upper limit of a bulk carrier density of $4 \times 10^{18}$ cm$^{-3}$ and, for this reason cannot be related with the bulk. We therefore associate $F_1$ to the bulk and $F_2$ to the TSS at the top surface.

For thick nanoribbons, the angular dependence of the magnetoresistance $R_{xx}$, which is used to map the Fermi surface, should indicate a 3D behaviour *i.e.* deviation from the $1/\cos(\theta)$ dependence. Fig. 4c illustrates the angular dependence of the SdH oscillations for a 60 nm nanoribbon. The oscillatory part $\Delta R_{xx}$ is plotted as a function of $1/B \cos(\theta)$. The oscillations are expected to align if they follow a $1/\cos(\theta)$ dependence. As one can see, a clear deviation is instead detected for angles $\theta > 40°$ (see Suppl. Info. S6 for the detailed extraction of the $1/\cos(\theta)$ dependence)

With the attribution of $F_1$ to the bulk we can calculate the corresponding bulk density $n_B$. The values we obtain for nanoribbons BR3-10R2 and B21-B1 ($t$ = 63 and 59 nm) are on the order of $2 \times 10^{18}$ cm$^{-3}$ which is in good agreement with the value $4 \times 10^{18}$ cm$^{-3}$ we have assumed as upper limit for the bulk contribution in thin nanoribbons. The interface carrier density $n_{Int}$ for the thick nanoribbons, with multi-frequency SdH oscillations, can be calculated from the total carrier density as $n_{Int} = n_{2D,H} - (n_B \times t + n_{TS})$. The value $n_{Int} = 3.4 \times 10^{13}$ cm$^{-2}$ is similar to the previous value derived from the two-carrier analysis of the thin nanoribbons. The overall results of magneto transport in thick nanoribbons confirms the high reproducibility of the properties of the TSS and interface states in our Bi$_2$Se$_3$ nanostructures.

It is also worth pointing out that the temperature dependence of $n_{3D}$ shown in Fig.1d cannot be attributed to a temperature dependence of bulk carriers. Indeed by assuming a temperature independent band bending the overall $n_{3D}$ should stay rather constant for a Fermi energy of 60 meV (Table 1) in the bulk conduction band. The temperature dependence of $n_{3D}$ can instead be attributed to an interface accumulation layer carrier concentration that depends on temperature, which itself can also depend on the nanobelt thickness. Such a dependence is not surprising since similar behavior has been observed in 2DEG gases at LAO/STO interfaces and appear to be a general property of interfaces between two oxide materials [36]. The physical origin of the accumulation layer at the interface with the substrate is possibly connected to the interface between the oxide layer surrounding the stoichiometric Bi$_2$Se$_3$ nanoribbons and the SiO$_2$/Si substrate (Suppl. Info., Fig. S12). It is well established that oxide interfaces may exhibit novel properties that are not found in the constituent materials [37,38]. A striking example is the LAO/STO interface [36]. In high-k/SiO$_2$ interfaces, for example, oxygen displacement at the interface is considered to be responsible for the formation of an interface dipole with an orientation depending on the areal density difference of oxygen atoms at the interface [39]. Depending on the sign of the dipole electrons an accumulation layer can be formed at the interface between the oxides to compensate the dipole electric field. In our case the oxide surrounding the nanoribbon and the SiO$_2$ are mainly amorphous, with some polycrystalline grains which could be associated with Bi$_2$O$_3$ and SeO$_2$, coupled through Van der Waals forces to the substrate (the nanoribbons are mechanically transferred to the substrate), (Suppl.

Info., Fig. S12). Only detailed atomistic calculations could therefore possibly give a clear insight into the effective species displacements at the interface accompanied by an electron density redistribution, leading to the formation of a 2DEG.

Finally the accumulation layer at the substrate interface can be removed by suspending the nanoribbons or by using different substrates and/or properly engineered buffer layer [20].

The presence of a depletion layer at the top surface (interfacing the vacuum) is a peculiarity of the PVD technique we have used to grow the nanoribbons. This is also demonstrated in other reports using a similar growth process [34], and contrasts with the finding of accumulation layers formed at the top surface of single crystals [31,40] and MBE thin films [27,41,42]. At the same time signatures of a TSS at the bottom interface, when an in plane magnetic field aligned with the nanoribbon axis is applied to the sample have been also detected in magnetotransport. In these measurements, for thin nanoribbon, we clearly see Aharonov-Bohm oscillations associated with orbits around the cross section (see Suppl. Info., Fig. S11). This indicates that despite the lower mobility of the bottom TSS, coherent trajectories are established on all 4 surfaces of the nanoribbon.

To conclude, our results highlight the promise of controlling the properties of TI materials through growth techniques and dimensionality and establish TI nanoribbon as a viable platform to study new phenomena and effects deriving from the topological protection of the surface states. Furthermore, it shows that post-growth treatment is not necessary to achieve bulk-free transport, significantly simplifying and lowering the requirements for eventual applications of these nanomaterials

**Methods:**

Transport measurements were performed in the Physical Property Measurement System (PPMS) by Quantum Design, equippe with a 14 T magnet and a base temperature of 2 K.

*Modelling details.* We solve self-consistently Poisson's equation for the electrostatic potential V:

$$\nabla^2 V(z) = e[n_{3D}(z) + n_{TSS}(z) + N_d]/\varepsilon,$$

where $\varepsilon$ is the dielectric constant of bulk Bi$_2$Se$_3$, $n_{3D}$ is the free-carrier density due to bulk and 2DEG bands, $n_{TS}$ is the carrier density due to the TSS and $N_d$ is the bulk dopant concentration. $n_{3D}$ is obtained by integrating the electron and hole density of states due to conduction and valence bands (assumed parabolic and using the effective mass approximation) up to the metal Fermi level $E_F$. $n_{TS}$ is obtained using the Dirac-like linear energy-dispersion of the TSS and assuming that they are spatially localized uniformly within 3 nm about each Bi$_2$Se$_3$ surface. The conduction and valence bands as well as the TSS are locally shifted by the self- local electrostatic potential within the conventional rigid shift approximation. We use $\varepsilon = 100$, which is representative of the available experimental values of the static dielectric constant for Bi$_2$Se$_3$[43], but note that more refined calculations may need to consider the anisotropy of the static dielectric constant with the direction of the electric field.

Solving Poisson's equation requires two boundary conditions, which we choose based on input from the experimental measurements. We set them by fixing $V(z)$ at the two ends of the simulation cell. For the simulations shown in Fig. 3a we impose that the CBM is 300 meV below $E_F$ at $z = 0$ and 10 meV above $E_F$ at $z = t$. The resulting average electronic charge density from these simulations: $\frac{1}{t}\int_0^t dz [n_{3D}(z) + n_{TSS}(z)]$ is shown in Fig. 3c as function of $t$. For the simulations shown in Supplementary Information Fig. S10 we choose $V(0)$ such that the electronic charge density integrated from 0 to t - 3 nm: $\int_0^{t-3nm} dz [n_{3D}(z) + n_{TSS}(z)]$ is within 1% of the value $n_{Int}$ shown in Table 1 while at the top surface we impose that the energy of the Dirac point equals the value $E_F^{TS}$ shown in Table 1.

**Acknowledgements:** The work has been supported by the Knut and Alice Wallenberg Foundation under the project "Dirac Materials" and the Latvian National Research Program IMIS 2. We also are grateful for support from the European Union for NANOCOHYBRI project (Cost Action CA 16218) and for the project 766714 – HiTIMe. GK has been partially financed by the Swedish Institute under the Visby project and the ERDF project 1.1.1.2/16/I/001, application No 1.1.1.2/VIAA/1/16/198. The support from the Swedish Infrastructure for Micro- and Nanofabrication- Myfab is acknowledged. The authors would like to thank Krisjanis Smits for help with the HR-TEM images.

**References:**

(1) Fu, L.; Kane, C. L. Superconducting Proximity Effect and Majorana Fermions at the Surface of a Topological Insulator. *Phys. Rev. Lett.* **2008**, *100*, 96407.

(2) Hasan, M. Z.; Kane, C. L. Colloquium: Topological Insulators. *Rev. Mod. Phys.* **2010**, *82*, 3045–3067.

(3) Pesin, D.; MacDonald, A. H. Spintronics and Pseudospintronics in Graphene and Topological Insulators. *Nat. Mater.* **2012**, *11*, 409–416.

(4) Mellnik, A. R.; Lee, J. S.; Richardella, A.; Grab, J. L.; Mintun, P. J.; Fischer, M. H.; Vaezi, A.; Manchon, A.; Kim, E.-A.; Samarth, N.; *et al.* Spin-Transfer Torque Generated by a Topological Insulator. *Nature* **2014**, *511*, 449–451.

(5) Shiomi, Y.; Nomura, K.; Kajiwara, Y.; Eto, K.; Novak, M.; Segawa, K.; Ando, Y.; Saitoh, E. Spin-Electricity Conversion Induced by Spin Injection into Topological Insulators. *Phys. Rev. Lett.* **2014**, 19660.

(6) Peng, H.; Lai, K.; Kong, D.; Meister, S.; Chen, Y.; Qi, X.-L.; Zhang, S.-C.; Shen, Z.-X.; Cui, Y. Aharonov-Bohm Interference in Topological Insulator Nanoribbons. *Nat. Mater.* **2010**, *9*, 225–229.

(7) Xiu, F.; He, L.; Wang, Y.; Cheng, L.; Chang, L.-T.; Lang, M.; Huang, G.; Kou, X.; Zhou, Y.; Jiang, X.; *et al.* Manipulating Surface States in Topological Insulator Nanoribbons. *Nat. Nanotechnol.* **2011**, *6*, 216–221.

(8) Hong, S. S.; Zhang, Y.; Cha, J. J.; Qi, X. L.; Cui, Y. One-Dimensional Helical Transport in Topological Insulator Nanowire Interferometers. *Nano Latters* **2014**, *14*, 2815–2821.

(9) Cho, S.; Dellabetta, B.; Zhong, R.; Schneeloch, J.; Liu, T.; Gu, G.; Gilbert, M. J.; Mason, N. Aharonov-Bohm Oscillations in a Quasi-Ballistic Three-Dimensional Topological Insulator Nanowire. *Nat. Commun.* **2015**, *6*, 7634.

(10) Jauregui, L. A.; Pettes, M. T.; Rokhinson, L. P.; Shi, L.; Chen, Y. P. Magnetic Field-Induced Helical Mode

and Topological Transitions in a Topological Insulator Nanoribbon. *Nat. Nanotechnol.* **2016**, *11*, 345–351.

(11) Appelbaum, I.; Drew, H. D.; Fuhrer, M. S. Proposal for a Topological Plasmon Spin Rectifier. *Appl. Phys. Lett.* **2011**, *98*, 023103–3.

(12) Tian, J.; Hong, S.; Miotkowski, I.; Datta, S.; Chen, Y. P. Observation of Current-Induced, Long-Lived Persistent Spin Polarization in a Topological Insulator: A Rechargeable Spin Battery. *Sci. Adv.* **2017**, *3*, e1602531.

(13) Cook, A.; Franz, M. Majorana Fermions in a Topological-Insulator Nanowire Proximity-Coupled to an S-Wave Superconductor. *Phys. Rev. B - Rapid Commun.* **2011**, *84*, 201105–4.

(14) Cook, A. M.; Vazifeh, M. M.; Franz, M. Stability of Majorana Fermions in Proximity-Coupled Topological Insulator Nanowires. *Phys. Rev. B* **2012**, *86*, 155431–17.

(15) de Juan, F.; Ilan, R.; Bardarson, J. H. Robust Transport Signatures of Topological Superconductivity in Topological Insulator Nanowires. *Phys. Rev. Lett.* **2014**, *113*, 107003.

(16) Ilan, R.; Bardarson, J. H.; Sim, H. S.; Moore, J. E. Detecting Perfect Transmission in Josephson Junctions on the Surface of Three Dimensional Topological Insulators. *New J. Phys.* **2014**, *16*, 053007–053013.

(17) Manousakis, J.; Altland, A.; Bagrets, D.; Egger, R.; Ando, Y. Majorana Qubits in a Topological Insulator Nanoribbon Architecture. *Phys. Rev. B* **2017**, *95*, 165424.

(18) Xue, L.; Zhou, P.; Zhang, C. X.; He, C. Y.; Hao, G. L.; Sun, L. Z.; Zhong, J. X. First-Principles Study of Native Point Defects in $Bi_2Se_3$. *AIP Adv.* **2013**, *3*, 52105.

(19) Xu, Y.; Miotkowski, I.; Liu, C.; Tian, J.; Nam, H.; Alidoust, N.; Hu, J.; Shih, C.-K.; Hasan, M. Z.; Chen, Y. P. Observation of Topological Surface State Quantum Hall Effect in an Intrinsic Three-Dimensional Topological Insulator. *Nat. Phys.* **2014**, *10*, 956–963.

(20) Koirala, N.; Brahlek, M.; Salehi, M.; Wu, L.; Dai, J.; Waugh, J.; Nummy, T.; Han, M. G.; Moon, J.; Zhu, Y.; *et al.* Record Surface State Mobility and Quantum Hall Effect in Topological Insulator Thin Films via Interface Engineering. *Nano Lett.* **2015**, *15*, 8245–8249.

(21) Moon, J.; Koirala, N.; Salehi, M.; Zhang, W.; Wu, W.; Oh, S. Solution to the Hole-Doping Problem and Tunable Quantum Hall Effect in $Bi_2Se_3$ Thin Films. *Nano Lett.* **2018**, 820–826.

(22) Hong, S. S.; Cha, J. J.; Kong, D.; Cui, Y. Ultra-Low Carrier Concentration and Surface-Dominant Transport in Antimony-Doped $Bi_2Se_3$ Topological Insulator Nanoribbons. *Nat. Commun.* **2012**, *3*, 757.

(23) Andzane, J.; Kunakova, G.; Charpentier, S.; Hrkac, V.; Kienle, L.; Baitimirova, M.; Bauch, T.; Lombardi, F.; Erts, D. Catalyst-Free Vapour-Solid Technique for Deposition of $Bi_2Te_3$ and $Bi_2Se_3$ Nanowires/nanobelts with Topological Insulator Properties. *Nanoscale* **2015**, *7*, 15935.

(24) Storm, K.; Halvardsson, F.; Heurlin, M.; Lindgren, D.; Gustafsson, A.; Wu, P. M.; Monemar, B.; Samuelson, L. Spatially Resolved Hall Effect Measurement in a Single Semiconductor Nanowire. *Nat. Nanotechnol.* **2012**, *7*, 718–722.

(25) Kim, Y. S.; Brahlek, M.; Bansal, N.; Edrey, E.; Kapilevich, G. a.; Iida, K.; Tanimura, M.; Horibe, Y.; Cheong, S.-W.; Oh, S. Thickness-Dependent Bulk Properties and Weak Antilocalization Effect in Topological Insulator $Bi_2Se_3$. *Phys. Rev. B* **2011**, *84*, 73109.

(26) Ando, Y. Topological Insulator Materials. *J. Phys. Soc. Japan* **2013**, *82*, 102001–102032.

(27) Bansal, N.; Kim, Y. S.; Brahlek, M.; Edrey, E.; Oh, S. Thickness-Independent Transport Channels in Topological Insulator $Bi_2Se_3$ Thin Films. *Phys. Rev. Lett.* **2012**, *109*, 116804.

(28) Ren, Z.; Taskin, A.; Sasaki, S.; Segawa, K.; Ando, Y. Large Bulk Resistivity and Surface Quantum Oscillations in the Topological Insulator $Bi_2Te_2Se$. *Phys. Rev. B* **2010**, *82*, 241306.

(29) Eto, K.; Ren, Z.; Taskin, A. A.; Segawa, K.; Ando, Y. Angular-Dependent Oscillations of the Magnetoresistance in $Bi_2Se_3$ due to the Three-Dimensional Bulk Fermi Surface. *Phys. Rev. B* **2010**, *81*, 195309.

(30) Brahlek, M.; Koirala, N.; Bansal, N.; Oh, S. Transport Properties of Topological Insulators: Band Bending, Bulk Metal-to-Insulator Transition, and Weak Anti-Localization. *Solid State Commun.* **2015**, *215*, 54–62.

(31) Analytis, J. G.; Chu, J.-H.; Chen, Y.; Corredor, F.; McDonald, R. D.; Shen, Z. X.; Fisher, I. R. Bulk Fermi Surface Coexistence with Dirac Surface State in $Bi_2Se_3$: A Comparison of Photoemission and Shubnikov–de Haas Measurements. *Phys. Rev. B* **2010**, *81*, 205407.

(32) Zhang, H.; Liu, C.-X.; Qi, X.-L.; Dai, X.; Fang, Z.; Zhang, S.-C. Topological Insulators in $Bi_2Se_3$, $Bi_2Te_3$ and $Sb_2Te_3$ with a Single Dirac Cone on the Surface. *Nat. Phys.* **2009**, *5*, 438–442.

(33) Xia, Y.; Qian, D.; Hsieh, D.; Wray, L.; Pal, A.; Lin, H.; Bansil, A.; Grauer, D.; Hor, Y. S.; Cava, R. J.; *et al.* Observation of a Large-Gap Topological-Insulator Class with a Single Dirac Cone on the Surface. *Nat. Phys.* **2009**, *5*, 398–402.

(34) Veyrat, L.; Iacovella, F.; Dufouleur, J.; Nowka, C.; Funke, H.; Yang, M.; Escoffier, W.; Goiran, M.; Eichler, B.; Schmidt, O. G.; *et al.* Band Bending Inversion in $Bi_2Se_3$ Nanostructures. *Nano Lett.* **2015**, *15*, 7503–7507.

(35) Sacépé, B.; Oostinga, J. B.; Li, J.; Ubaldini, A.; Couto, N. J. G.; Giannini, E.; Morpurgo, A. F. Gate-Tuned Normal and Superconducting Transport at the Surface of a Topological Insulator. *Nat. Commun.* **2011**, *2*, 575.

(36) Huijben, M.; Koster, G.; Kruize, M. K.; Wenderich, S.; Verbeeck, J.; Bals, S.; Slooten, E.; Shi, B.; Molegraaf, H. J. A.; Kleibeuker, J. E.; *et al.* Defect Engineering in Oxide Heterostructures by Enhanced Oxygen Surface Exchange. *Adv. Funct. Mater.* **2013**, *23*, 5240–5248.

(37) Park, J. W.; Bogorin, D. F.; Cen, C.; Felker, D. A.; Zhang, Y.; Nelson, C. T.; Bark, C. W.; Folkman, C. M.; Pan, X. Q.; Rzchowski, M. S.; *et al.* Creation of a Two-Dimensional Electron Gas at an Oxide Interface on Silicon. *Nat. Commun.* **2010**, *1*, 94.

(38) Hwang, H. Y.; Iwasa, Y.; Kawasaki, M.; Keimer, B.; Nagaosa, N.; Tokura, Y. Emergent Phenomena at Oxide Interfaces. *Nat. Mater.* **2012**, *11*, 103–113.

(39) Kita, K.; Toriumi, A. Intrinsic Origin of Electric Dipoles Formed at High-k/$SiO_2$ Interface. *Appl. Phys. Lett.* **2009**, 132902.

(40) Bianchi, M.; Guan, D.; Bao, S.; Mi, J.; Iversen, B. B.; King, P. D. C.; Hofmann, P. Coexistence of the Topological State and a Two-Dimensional Electron Gas on the Surface of $Bi_2Se_3$. *Nat. Commun.* **2010**, *1*, 125–128.

(41) Brahlek, M.; Kim, Y. S.; Bansal, N.; Edrey, E.; Oh, S. Surface versus Bulk State in Topological Insulator $Bi_2Se_3$ under Environmental Disorder. *Appl. Phys. Lett.* **2011**, *99*, 10–13.

(42) Brahlek, M.; Koirala, N.; Salehi, M.; Bansal, N.; Oh, S. Emergence of Decoupled Surface Transport Channels

in Bulk Insulating $Bi_2Se_3$ Thin Films. *Phys. Rev. Lett.* **2014**, *113*, 1–5.

(43) Butch, N. P.; Kirshenbaum, K.; Syers, P.; Sushkov, a. B.; Jenkins, G. S.; Drew, H. D.; Paglione, J. Strong Surface Scattering in Ultrahigh-Mobility $Bi_2Se_3$ Topological Insulator Crystals. *Phys. Rev. B* **2010**, *81*, 241301.

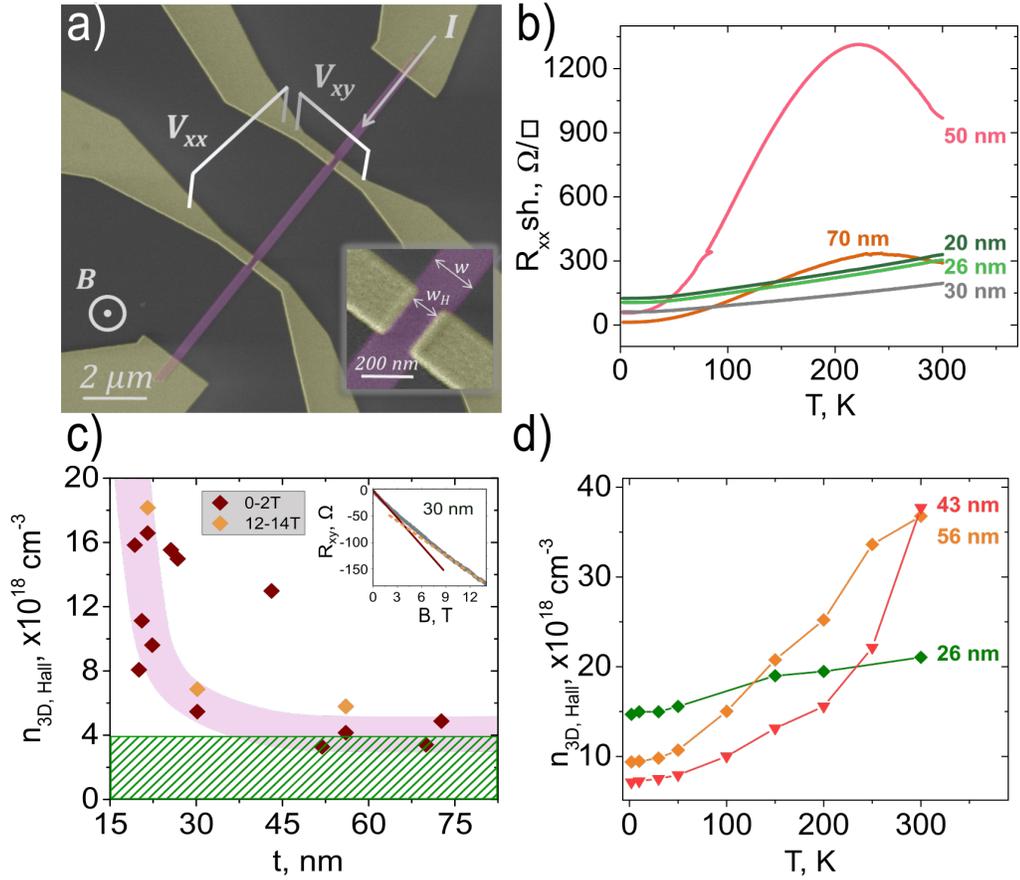

**FIG. 1.** a) Coloured SEM image (electrodes in yellow and Bi$_2$Se$_3$ nanoribbon in violet) of an individual Bi$_2$Se$_3$ nanoribbon with 6 contacts enabling measurements of the longitudinal ($V_{xx}$) and transverse ($V_{xy}$) voltages. b) Temperature dependence of the longitudinal sheet resistance of Bi$_2$Se$_3$ nanoribbons with different thicknesses. c) Calculated 3D carrier density extracted from the Hall effect in the magnetic field range 0 – 2 T (red diamonds) as a function of the nanoribbon thickness; the orange diamonds represent $n_{3D}$ extracted at high magnetic field (12 – 14 T); T=2K. The inset shows the magnetic field dependence of the transverse resistance $R_{xy}$ for a nanoribbon with $t$ = 30 nm (red solid and orange dashed curves represent linear fits to magnetic field range 0 – 2 T and 12 – 14 T). The pink shaded region is a guide to the eye and the green striped region indicates the upper bound for the bulk carrier concentration. d) Temperature dependence of the 3D carrier density $n_{3D}$ extracted from the Hall effect in the magnetic field range 0 – 2 T for nanoribbons with different thicknesses.
17

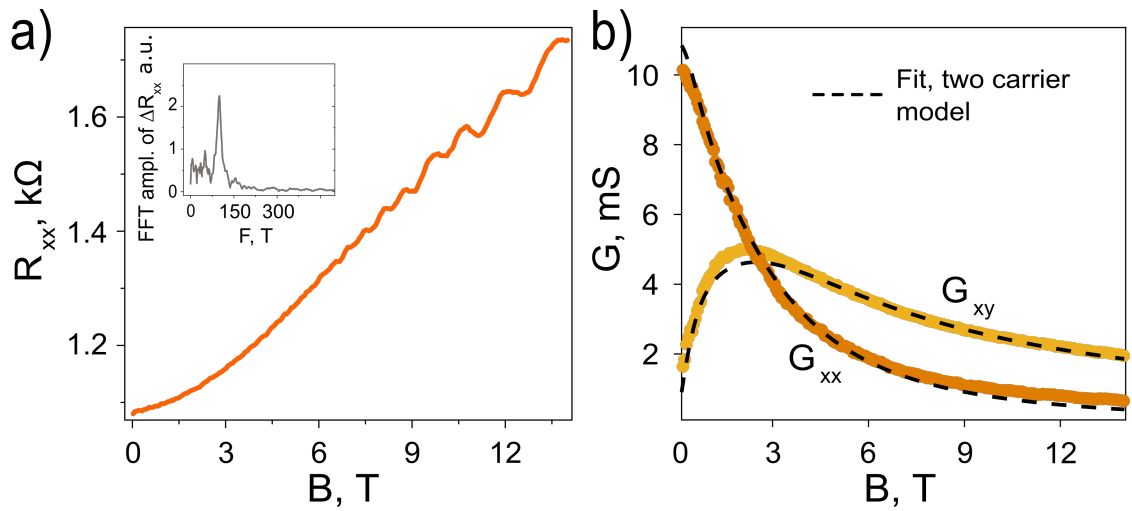

**FIG. 2.** a) Shubnikov-de Haas oscillations for a $Bi_2Se_3$ nanoribbon (device B13-E5). The inset shows the FFT spectrum of $\Delta R_{xx}$. A single dominating pick is observed at a frequency F = 99 T. An additional peak might be present at lower frequency; however, the signal to noise ratio is too low to draw firm conclusions. b) Longitudinal and transverse conductance versus magnetic field. The dashed lines correspond to the fit of the two-carrier model.

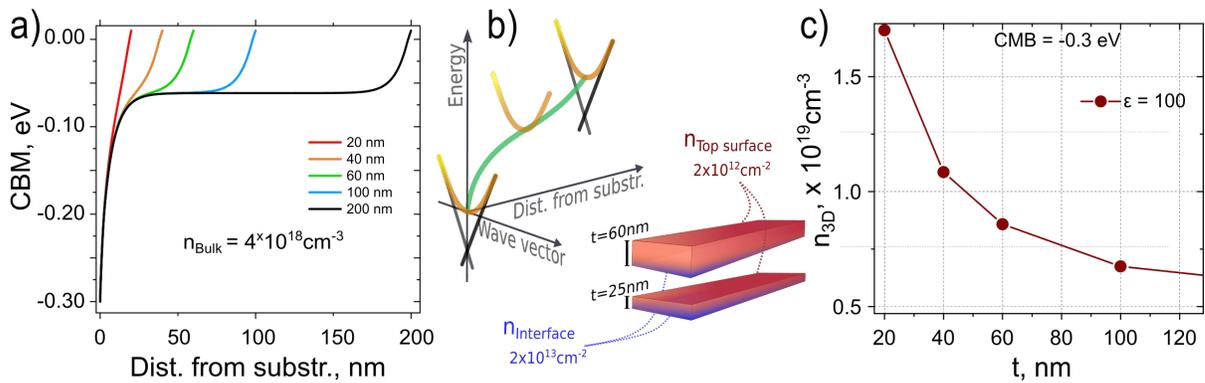

**FIG. 3** a) Calculated CBM as a function of the distance from the substrate for nanoribbons of different thicknesses. b) 3D Schematics: on the left – the band-bending for a 30 nm nanoribbon; on the right – 25 and 60 nm thick nanoribbons c) Calculated effective carrier density from the self-consistent band-bending of panel a, using a dielectric constant ε =100.

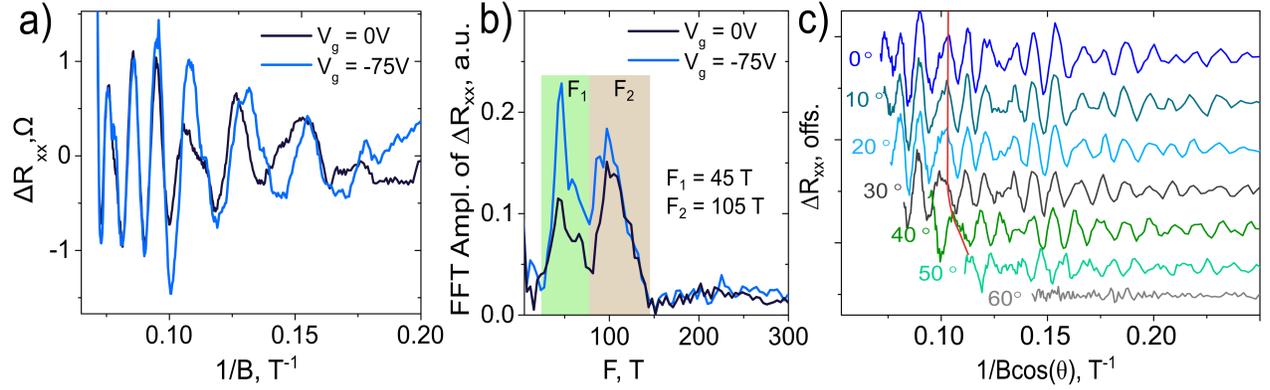

**FIG.4.** a) Shubnikov-de Haas oscillations of a $Bi_2Se_3$ nanoribbon with thickness 63 nm at 0 and − 75 V back gate voltages and b) FFT spectra of SdH oscillations of a) with a removed background $\Delta R_{xx}$. c) $\Delta R_{xx}$ *versus* $1/B \cos(\theta)$ for various angles $\theta$ between the magnetic field and the surface normal. The red line is a guide to the eye showing the departure from the $1/B \cos(\theta)$ scaling of the SdH oscillations.

**TABLE 1.** Summary of characteristic parameters of $Bi_2Se_3$ nanoribbons with different thicknesses.[*]

| Nr | $t$ (nm) | $n_{TS}$ (cm$^{-2}$) | $n_{Int}$ (cm$^{-2}$) | $n_B$ (cm$^{-3}$) | $E_F^{TS}$ (meV) | $E_F^B$ (meV) | $\Delta E_{BB}^{TS}$ (meV) | $\Delta E_{BB,}^{Int}$ (meV) |
|---|---|---|---|---|---|---|---|---|
| B13-E3 | 26 | $2.1\times10^{12}$ | $3.8\times10^{13}$ [a] | | 170 | | 71 | −280 |
| B13-C3 | 21 | $2.2\times10^{12}$ | $1.9\times10^{13}$ [a] | $4\times10^{18}$ | 172 | 60 | 68 | −130 |
| B13-E5 | 30 | $2.4\times10^{12}$ | $1.5\times10^{13}$ [b] | | 180 | | 61 | −40 |
| BR3-10R2 | 63 | $2.5\times10^{12}$ | - | $1.7\times10^{18}$ | 185 | 35 | 29 | - |
| B21-B1 | 59 | $2.6\times10^{12}$ | $3.4\times10^{13}$ | $2.2\times10^{18}$ | 187 | 41 | 34 | −200 |

[*]The band-bending energy at the interface with the substrate $\Delta E_{BB,}^{Int}$ is calculated as $\Delta E_{BB,}^{Int} = E_F^B - CBM(0)$, where the conduction band minimum at the interface with the substrate is extracted from the simulated band-bending diagrams (Suppl. Info, Fig. S10).

[a] calculated as $n_{Int} = n_{2D,H} - n_{TS}$

[b] Determined from two – carrier analysis

# Supplementary Information

# Bulk-Free Topological Insulator $Bi_2Se_3$ nanoribbons with Magnetotransport Signatures of Dirac Surface States

Gunta Kunakova[1,2], Luca Galletti[1], Sophie Charpentier[1], Jana Andzane[2], Donats Erts[2], François Léonard[3], Catalin D. Spataru[3], Thilo Bauch[1] and Floriana Lombardi[1]

[1])*Quantum Device Physics Laboratory, Department of Microtechnology and Nanoscience, Chalmers University of Technology, SE-41296 Gothenburg, Sweden*
[2])*Institute of Chemical Physics, University of Latvia, Raina Blvd. 19, LV-1586, Riga, Latvia*
[3])*Sandia National Laboratories, Livermore, CA, 94551, United States*

## 1) Shubnikov-de Haas oscillations in Bi$_2$Se$_3$ nanoribbons

*1-1) Thin nanoribbons (t<35nm)*

In Fig. S1-a we show Shubnikov-de Haas (SdH) oscillations for three nanoribbons with thicknesses 26 – 31 nm. The Fast Fourier Transform (FFT) spectra give one dominating frequency (*F*) around 100 T (Fig. S1-b). The determined frequency of SdH oscillations is used to calculate the 2D carrier density $n_{2D,SdH} = k^2_F/4\pi = 2\times10^{12}$ cm$^{-2}$ (Table S1).

The cyclotron mass $m_c$ is calculated in agreement with the Lifshitz – Kosevich theory. The $\Delta R_{xx}$ values as a function of temperature at fixed magnetic field positions corresponding to the Landau Indices *N* = 7; 8; 9 are shown in Fig. S2-a. The amplitude of the oscillations with the increase of the temperature can be expressed as $\Delta R_{xx}(T) \propto (2\pi^2 k_B T / \hbar\omega) / (\sinh(2\pi^2 k_B T / \hbar\omega))$, where $k_B$ is the Boltzmann constant, $\hbar$ is the reduced Planck constant, $\omega$ is the cyclotron frequency[1]. By fitting the temperature dependent amplitude of the SdH oscillations in agreement with this equation, we extract $\omega$, which is then used to calculate the cyclotron mass as $m_c = eB/\omega$. The cyclotron mass is determined for 5 different magnetic field positions and the averaged $m_c$ values are listed in the Table S1.

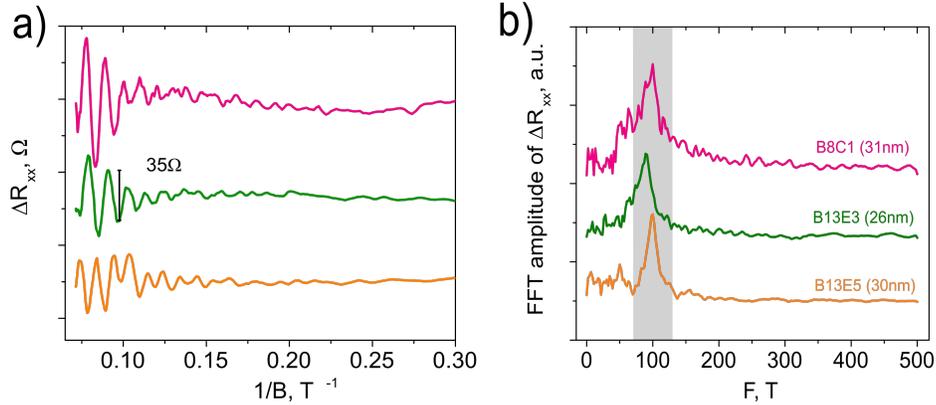

FIG. S1. a) Oscillatory part of the longitudinal magnetoresistance $\Delta R_{xx}$ with removed background as a function of 1/*B*. b) FFT amplitude of the $\Delta R_{xx}$ for nanoribbons with thicknesses 26 – 31 nm.

To evaluate the 2D carrier mobility we performed a Dingle analysis (Fig. S2-b). Here we extracted the quantum lifetime $\tau$ according to the relation $\ln(\Delta R_{xx} B \sinh(\lambda(T))) \approx 1/B \times (2\pi^2 E_F/\tau e v_F^2)$, where the thermal factor $\lambda = 2\pi^2 k_B T m_c/eB$. Extracted $\tau$ was then used to calculate the 2D carrier mobility $\mu_{SdH} = e\tau/m_c$, where $m_c$ is the cyclotron mass. Determined values of the $\mu_{SdH}$ are given in Table S1.

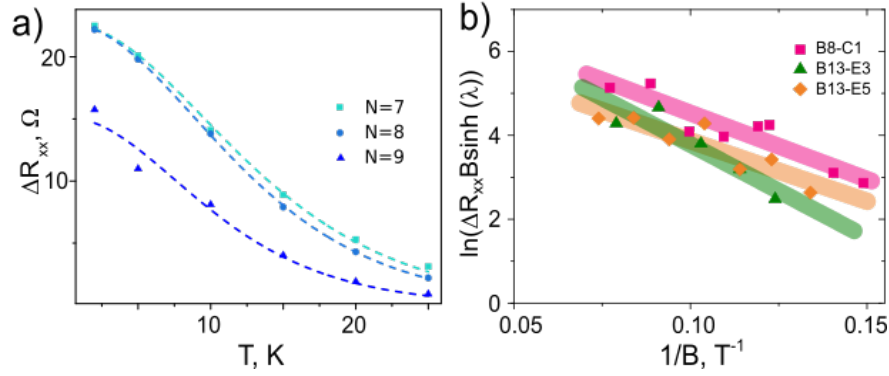

FIG. S2. a) Amplitude $\Delta R_{xx}$ of the SdH oscillations as a function of temperature for nanoribbon B13-E3. The dashed curves correspond to the fit used to extract the cyclotron frequency. N = 7; 8; 9 are assigned Landau indexes for several oscillation peaks. b) Dingle plots of three nanoribbons listed in Table S1. Solid – transparent lines are the fits used to extract the quantum lifetime and the corresponding 2D mobility.

TABLE S1. Summary of the calculated parameters from the SdH oscillations for the $Bi_2Se_3$ nanoribbons.

| No | Nanoribbon | $t$, nm | $F$, T | $n_{2D,SdH}$, $cm^{-2}$ | $\beta$ | $\mu_{SdH}$, $cm^2(Vs)^{-1}$ | $m_c / m_e$ |
|---|---|---|---|---|---|---|---|
| 1 | B13-E3 | 26 | 88 | $2.1\times10^{12}$ | 0.48±0.03 | 4500±1100 | 0.145±0.021* |
| 2 | B13-E5 | 30 | 99 | $2.4\times10^{12}$ | 0.23±0.06 | 6800±1400 | * |
| 3 | B8-C1 | 31 | 96 | $2.3\times10^{12}$ | 0.22±0.03 | 6200±900 | 0.136±0.011 |

*The value of cyclotron mass of nanoribbon B13-E3 is used in the $\mu_{SdH}$ calculations for nanoribbon B13-E5.

SdH oscillations for thin nanoribbons (t = 26 – 31 nm) are represented by a single oscillation frequency. We can use these data to construct a Landau level fan diagrams for extracting a Berry phase (this is not the case for complicated multi-frequency SdH oscillation patterns, as observed for thicker nanoribbons, where reliable fan diagrams cannot be constructed due to overlapping of several frequencies). The Berry phase can be determined from a linear fit of Landau level fan diagram. Here the intercept $\beta$ with the $y$ axis relates with the Berry phase as $\beta = \varphi/2\pi$. For Dirac fermions, the Berry phase $\varphi = \pi$ and the intercept $\beta = 0.5$.

A reliable Landau level fan diagram can be constructed considering both longitudinal and transversal components of the magnetotransport measurements, i.e., from conductivity tensor $\sigma_{xx} = \frac{\rho_{xx}}{\rho_{xx}^2+\rho_{xy}^2}$; $\sigma_{xy} = -\frac{\rho_{xy}}{\rho_{xy}^2+\rho_{xx}^2}$ and the Landau levels are indexed by taking the minima of $\sigma_{xx}^2$. Depending on $\rho_{xx}$ and $\rho_{xy}$ ratio, there are two different conditions [2]:

1) $\sigma_{xx} \ll \sigma_{xy}$, $\rho_{xx} \approx \sigma_{xx}/\sigma_{xy}^2$ ; *min* in $\rho_{xx}$ is *min* in $\sigma_{xx}$

2) $\sigma_{xx} \gg \sigma_{xy}$, $\rho_{xx} \approx 1/\sigma_{xx}$ ; *max* in $\rho_{xx}$ is *min* in $\sigma_{xx}$.

By converting magnetoresistance data to conductivity tensor, we find that the $\sigma_{xx} < \sigma_{xy}$ (see also Fig. 2b in the main text, where the conductance tensor data is plotted for the two-carrier analysis) and *min* in $\rho_{xx}$ should coincide with *min* in $\sigma_{xx}$.

Fig. S3-a shows SdH oscillations for nanoribbon B13-E5, thickness – 30nm. Both $\Delta\sigma_{xx}$ – tensor and plain $\Delta\sigma_{xx}$ (=1/$\rho_{xx}$) data with removed polynomial background are plotted. As expected for $\sigma_{xx} < \sigma_{xy}$ condition, minima in $\sigma_{xx}$ – tensor occur at the maxima of plain $\sigma_{xx}$ (or at minima of $\rho_{xx}$). The SdH oscillations in $\sigma_{xx}$ – tensor have with a lower amplitude and are noisy compared to plain $\sigma_{xx}$ due to the impact of low signal to noise $R_{xy}$ data. Therefore, the best way to construct the Landau level diagrams for our $Bi_2Se_3$ nanoribbons would be to index the minima in $\rho_{xx}$. The constructed Landau level fan diagrams with a linear fit for all three thin $Bi_2Se_3$ nanoribbons are shown in Fig. S3-b. Extracted intercepts $\beta$ = 0.22 – 0.48 are summarized in Table S1.

To validate if the linear fitting in the Landau level fan diagrams is performed correctly, we plot the residual $\Delta 1/B$ values from linear fit as a function of Landau indices N. As one can see, the residual of linear fit is scattered around zero for all three nanoribbons (Fig. S3-c).

The deviation from the value 0.5 for the extrapolated intercept expected for a linear Dirac dispersion is most probably caused by a parabolic contribution to the Dirac cone, which shifts the observed Berry phase extracted from SdH oscillation towards zero [3]. Such a parabolic contribution to the Dirac cone is consistently observed in ARPES measurements of $Bi_2Se_3$ crystals.

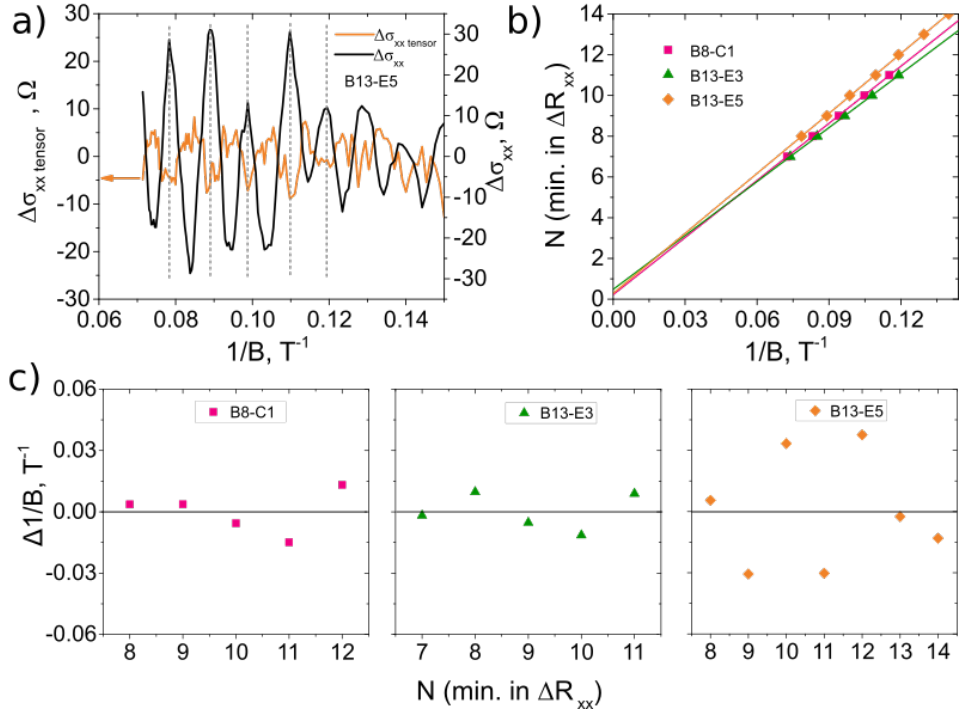

FIG. S3. Extraction of Berry phase for thin $Bi_2Se_3$ nanoribbons. a) $\sigma_{xx}$ – tensor and plain $\sigma_{xx}(=1/\rho_{xx})$ data with removed polynomial background plotted as a function of 1/B. b) Landau level diagrams of three nanoribbons listed in the Table S1. The solid lines are fit to extract the intercept with the *y* axis. c) Residual values of Δ1/B from linear fit of Landau level diagrams versus Landau indices N labelled to minima of $\Delta R_{xx}$.

The angular dependence of the SdH oscillations for thin nanoribbon (B8-C1) is plotted in Fig. S4. For a 2D surface, the oscillation frequency is expected to follow F ∝ 1/ cos(θ), where θ is the angle between the top surface of the nanoribbon and the field direction. The fact, that SdH oscillations align for all angles if plotting $\Delta R_{xx}$ versus 1/B cos(θ), confirms the 2D nature of the SdH oscillations.

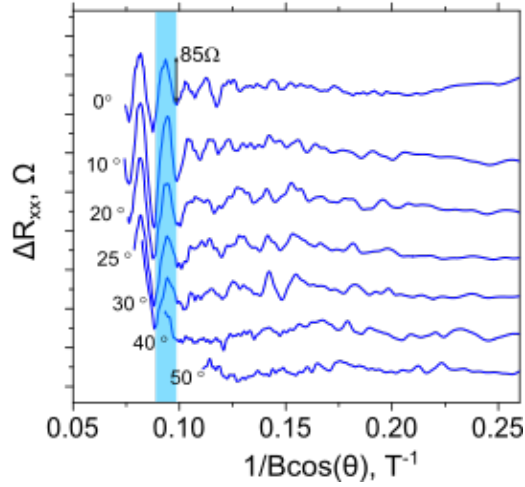

FIG. S4. SdH oscillations for the nanoribbon B8-C1 as a function of $1/B \cos(\theta)$. Angle at 0 deg represents magnetic field direction that is perpendicular to the nanoribbon top surface.

*1-2) Thick nanoribbons (t>35nm)*

Thicker $Bi_2Se_3$ nanoribbons (with thicknesses above 35nm) always shows multi-frequency SdH oscillation pattern. In Fig. S5 we plot the longitudinal magnetoresistance $\Delta R_{xx}$ for 59 nm thick nanoribbon B21-B1, (extracted carrier densities for this nanoribbon are included in the main text, Table1). The FFT spectrum gives two dominating frequencies, which correspond to the surface and bulk carriers as discussed in the main text.

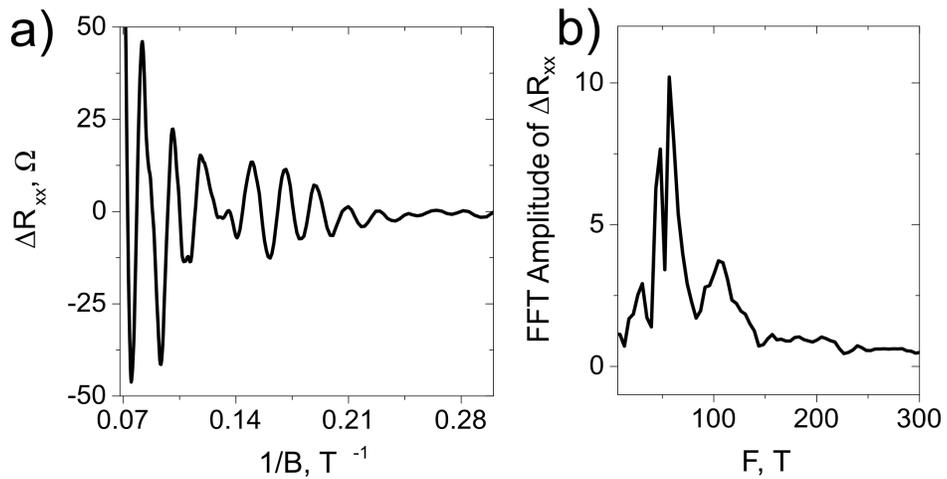

FIG. S5. a) Oscillatory part of the longitudinal magnetoresistance with removed smooth background $\Delta R_{xx}$ as a function of $1/B$. b) FFT of the $\Delta R_{xx}$. Data correspond to nanoribbon B21-B1, $t = 59$ nm.

The FFT spectra of SdH oscillations for various magnetic field angles with respect to the surface normal for a thicker nanoribbon, shown in Fig. 4c of the main text, is depicted in Fig. S6.

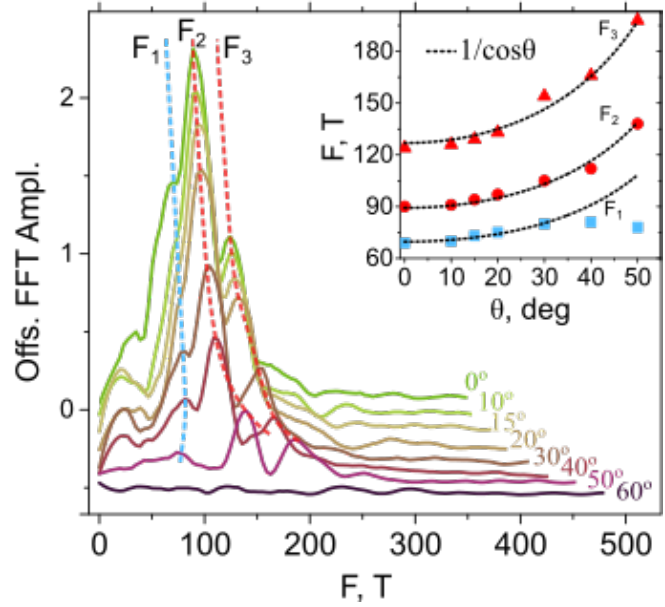

FIG. S6. FFT spectra of a thick TI nanoribbon ($t = 57$ nm) for various angles of the applied magnetic field θ, where θ=0° corresponds to a field applied perpendicular to the TI nanoribbon surface. The inset shows the evolution of the three main peaks as a function of angle θ.

The FFT spectra of this nanoribbon gives three frequencies: $F_1$, $F_2$ and $F_3$. Two higher frequencies $F_3 = 127$T and $F_2 = 90$T follows $1/\cos\theta$ dependence and they clearly originate from a 2D Fermi surface. These frequencies represent topological surface states at the nanoribbon top surface (see discussion in main text). The fact that there are two 2 oscillation frequencies with the $1/\cos\theta$ dependence can be related to existing terrace in the nanoribbon top surface (see AFM image of this nanoribbon in Fig. S7). Corresponding 2D carrier density calculated from $F_2$ and $F_3$ is 2.2 and $3\times10^{12}$cm$^{-2}$, which is similar with those carrier density values extracted from a single-frequency SdH oscillations (see in Table S1).

The lowest $F_1 = 70$T shows deviation from $1/\cos\theta$ dependence and is attributed to the bulk electrons. The bulk carrier density is calculated as $n_{3D} = 1/(2\pi)^2(4/3)\,k_F^3$, and $n_{3D}=3.3\times10^{18}$ cm$^{-3}$. This value is in agreement with an expected ellipsoidal Fermi surface for the bulk [4,5], resulting in a departure from the expected $1/\cos\theta$ dependence of SdH oscillation frequency for a 2D Fermi surface or a 3D cylindrical Fermi surface (limit of large bulk carrier concentration $10^{20}$ cm$^{-3}$).

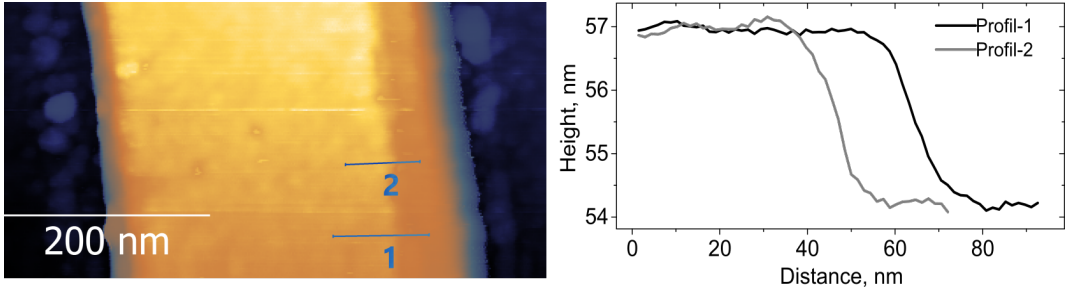

FIG. S7. Atomic Force Microscopy image of the $Bi_2Se_3$ nanoribbon shown in Fig. S6. A step/terraces can be clearly observed in the nanoribbon surface. The right panel shows line scans indicated by the lines 1 and 2 in the left panel.

## 2) Hall effect in individual nanoribbons

To extract the total Hall carrier density (and to perform a two-carrier analysis as discussed in the main text), we have measured $R_{xy}$ magnetoresistance at high magnetic field intensities up to 14 T. $R_{xy}$ characteristics are nonlinear at magnetic field values above 2T. This fact confirms multi-band transport in our TI nanoribbons. The total Hall carrier density can be determined from the high-field region 12 – 14T. The determined values are about 20% higher compare to those extracted from 0-2T region. These data are shown in the main text, Fig. 1c.

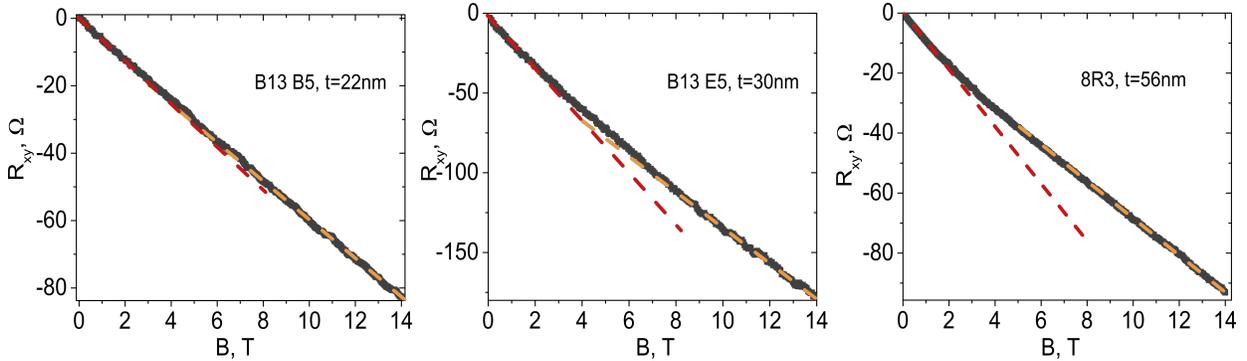

FIG. S8. Transversal resistance $R_{xy}$ as a function of magnetic field for $Bi_2Se_3$ nanoribbons with different thicknesses (T = 2K).

## 3) Gate dependence of $R_{xx}$ in individual nanoribbons

Fig S9 depicts gate dependence of the longitudinal resistance $R_{xx}$ at temperature of 2 K for nanoribbons with different thicknesses. The 3D TIs are ambipolar materials and the gate dependence of the $R_{xx}$ can give additional information about the charge neutrality point (CNP), which usually is observed at large negative voltages in case of $SiO_2$ as a gate dielectric[6,7].

The CNP can be reached for a thinner $Bi_2Se_3$ nanoribbon. As shown in Fig. S9-a, the CNP of the top surface (more than tenfold increase of $R_{xx}$) is reached at a back gate voltage of -70 V. The main reason for applying such high gate voltage is to compensate the accumulation layer at the TI interface with the substrate.

For the 65 nm thick nanoribbon we do not reach the CNP even for applied voltages of -90V (Fig. S9-b), which is compatible with the existence of bulk charge carriers in parallel with the surface states at the top TI surface and the accumulation layer at the TI interface with the substrate. The fact that in thick nanoribbons we have bulk carriers makes it close to impossible to reach the charge neutrality point at the TI top surface.

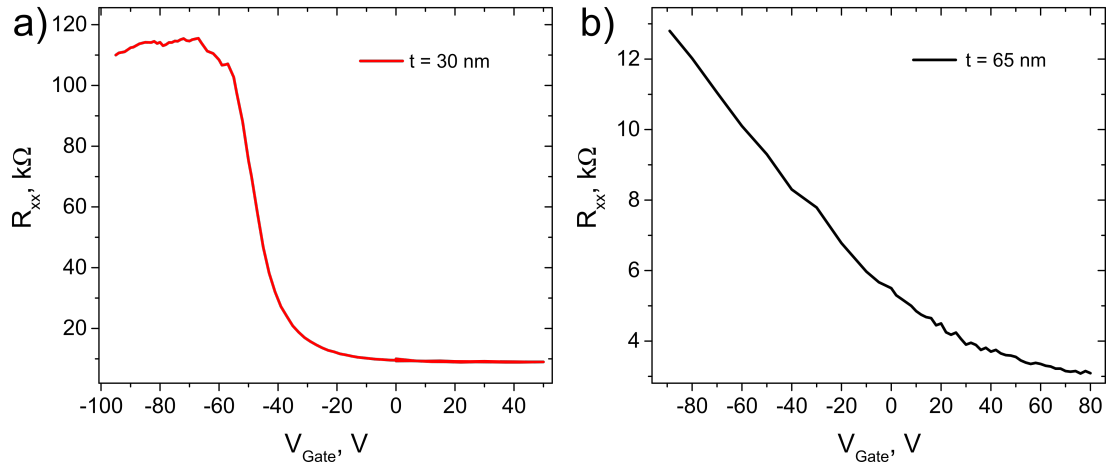

FIG. S9. Back - gate dependence of the longitudinal resistance at 2K for $Bi_2Se_3$ nanoribbons: a) t = 30nm and b) t = 65nm.

**4) Band bending for $Bi_2Se_3$ nanoribbons**

The band-bending diagrams for $Bi_2Se_3$ nanoribbons are modelled by solving the Poisson`s equation. For these simulations a self-consistent calculation for a slab of $Bi_2Se_3$ with the boundary conditions set by the experimentally determined values of different carrier populations in $Bi_2Se_3$ nanoribbons. Fig. S10 shows the simulated band-bending diagrams for thin $Bi_2Se_3$ nanoribbons using this approach. The conduction band minimum (CBM) at distance "0 nm" (the "Dist." in the plots is the distance of $Bi_2Se_3$ from the substrate) corresponds to the interface $Bi_2Se_3/SiO_2$, with an accumulation layer. The CMB values at Dist.=0 nm are used to evaluate the band-bending energy at the interface $\Delta E_{BB,}^{Int}$ (Table 1, main text).

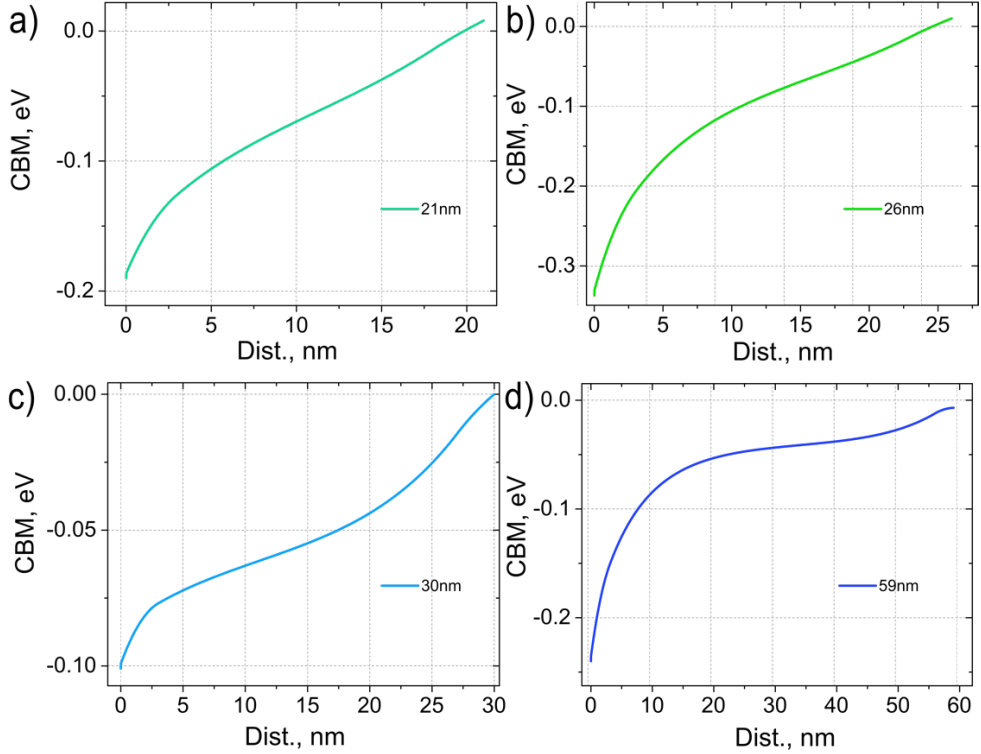

FIG. S10. a) – d) Band-bending diagrams for nanoribbons B13-C3 ($t$ = 21 nm); B13-E3 ($t$ = 26 nm); B13-E5 ($t$ = 30 nm) and B21-B1 ($t$ = 59 nm). The boundary conditions are set by the experimental values shown in the main text, Table 1. Dielectric constant used in the simulations $\varepsilon$ = 100.

## 5) Aharonov – Bohm oscillations

The longitudinal magnetoresistance of the $Bi_2Se_3$ nanoribbon B8-C1 ($t$ = 31 nm, $w$ = 260 nm) as a function of *axial* magnetic field is shown in Fig. S11. The $R_{xx}$ oscillates with the period of $\Delta B = \Phi_0/A$, where $\Phi_0 = h/e$ is the magnetic flux quantum and $A$ is the cross-section of a nanoribbon [8].

The cross-section area is calculated as $A = t \times w = 8.06 \times 10^{-15}$ m$^2$. This value of cross-section would correspond to the oscillation period of 0.51 T, which is different than the experimentally observed $\Delta B = 0.79$ T (Fig. S11). This discrepancy suggests, that the effective cross-section (the one representing the area where the charge transport takes place) is smaller than the geometrical. If we account for ~ 5 nm thick layer of a native oxide commonly present on the surfaces of $Bi_2Se_3$ and $Bi_2Te_3$[9], the cross-section area is $5.25 \times 10^{-15}$ m$^2$ matching the experimentally determined $\Delta B$ of 0.79 T.

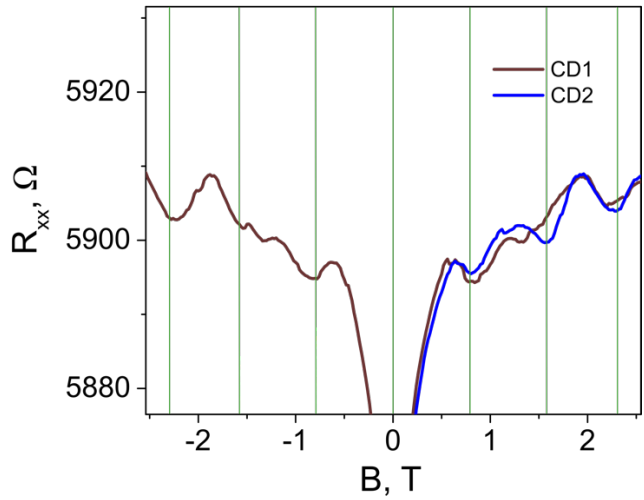

FIG. S11. Longitudinal resistance versus axial magnetic field. Data correspond to two thermal cycles CD1 and CD2 of the nanoribbon B8- C1 ($t$ = 31 nm, $w$ = 260 nm).

## 6) HR-TEM studies of the Bi$_2$Se$_3$ nanoribbons

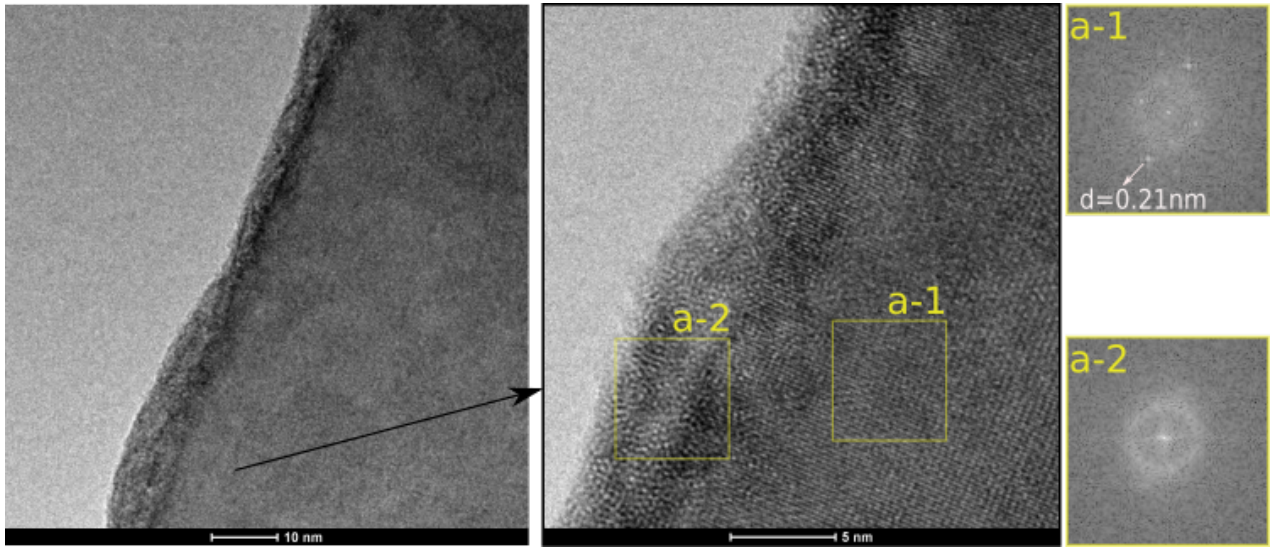

FIG. S12. High-resolution TEM images on a Bi$_2$Se$_3$ nanoribbon

Fig. S12 illustrates HR-TEM studies of a Bi$_2$Se$_3$ nanoribbon. The left panel provides an overview of a Bi$_2$Se$_3$ nanoribbon, showing an oxide layer covering the nanoribbon surface. Close view of this oxide layer (of an average thickness of ~5 nm) and its interface with the single-crystalline core of the nanoribbon is shown in higher magnification image (middle panel). The two panels on the right show FFT images corresponding to two different regions of the Bi$_2$Se$_3$ nanoribbon, labelled a-1 and a-2. The FFT image corresponding to the a-1 area (the inner structure of the nanoribbon) shows a hexagonally symmetric pattern, confirming single-crystalline structure of $Bi_2Se_3$. The interplane spacing extracted from this pattern is ~0.21 nm, which is consistent with the interplane distance of $[11\bar{2}0]$ of $Bi_2Se_3$[10]. The FFT image corresponding to the area a-2 (the surface layer of the nanoribbon) shows a ring indicating amorphous structure. The ring diameter approximately equates to a lattice spacing of 0.32 nm, meaning that the surface layer could correspond to a bismuth oxide[11] and/or selenium oxide[12] layer formed from air exposure.

**References:**

(1) Isihara, A.; Smrčka, L. Density and Magnetic Field Dependences of the Conductivity of Two-Dimensional Electron Systems. *J. Phys. C Solid State Phys.* **1986**, *19*, 6777–6789.

(2) Ando, Y. Topological Insulator Materials. *J. Phys. Soc. Japan* **2013**, *82*, 102001–102032.

(3) Taskin, A. A.; Ando, Y. Berry Phase of Nonideal Dirac Fermions in Topological Insulators. *Phys. Rev. B - Condens. Matter Mater. Phys.* **2011**, *84*, 1–6.

(4) Lawson, B. J.; Li, G.; Yu, F.; Asaba, T.; Tinsman, C.; Gao, T.; Wang, W.; Hor, Y. S.; Li, L. Quantum Oscillations in $Cu_xBi_2Se_3$ in High Magnetic Fields. *Phys. Rev. B* **2014**, *90*, 195141.

(5) Lahoud, E.; Maniv, E.; Petrushevsky, M. S.; Naamneh, M.; Ribak, A.; Wiedmann, S.; Petaccia, L.; Salman, Z.; Chashka, K. B.; Dagan, Y.; *et al.* Evolution of the Fermi Surface of a Doped Topological Insulator with Carrier Concentration. *Phys. Rev. B* **2013**, *88*, 195107.

(6) Checkelsky, J. G.; Hor, Y. S.; Cava, R. J.; Ong, N. P. Bulk Band Gap and Surface State Conduction Observed in Voltage-Tuned Crystals of the Topological Insulator $Bi_2Se_3$. *Phys. Rev. Lett.* **2011**, *106*, 196801.

(7) Hong, S. S.; Cha, J. J.; Kong, D.; Cui, Y. Ultra-Low Carrier Concentration and Surface-Dominant Transport in Antimony-Doped $Bi_2Se_3$ Topological Insulator Nanoribbons. *Nat. Commun.* **2012**, *3*, 757.

(8) Peng, H.; Lai, K.; Kong, D.; Meister, S.; Chen, Y.; Qi, X.-L.; Zhang, S.-C.; Shen, Z.-X.; Cui, Y. Aharonov-Bohm Interference in Topological Insulator Nanoribbons. *Nat. Mater.* **2010**, *9*, 225–229.

(9) Tian, M.; Ning, W.; Qu, Z.; Du, H.; Wang, J.; Zhang, Y. Dual Evidence of Surface Dirac States in Thin Cylindrical Topological Insulator $Bi_2Te_3$ Nanowires. *Sci. Rep.* **2013**, *3*, 1212.

(10) Liu, F.; Liu, M.; Liu, A.; Yang, C.; Chen, C.; Zhang, C.; Bi, D.; Man, B. The Effect of Temperature on $Bi_2Se_3$ Nanostructures Synthesized via Chemical Vapor Deposition. *J. Mater. Sci. Mater. Electron.* **2015**, *26*, 3881–3886.

(11) Friedensen, S.; Mlack, J. T.; Drndić, M. Materials Analysis and Focused Ion Beam Nanofabrication of Topological Insulator $Bi_2Se_3$. *Sci. Rep.* **2017**, *7*, 13466.

(12) Zhao, Q.; Zhang, H. Z.; Xiang, B.; Luo, X. H.; Sun, X. C.; Yu, D. P. Fabrication and Microstructure Analysis of SeO$_2$ Nanowires. *Appl. Phys. A Mater. Sci. Process.* **2004**, *79*, 2033–2036.